\documentclass[twocolumn, final, 10pt]{IEEEtran}
\usepackage[dvips]{graphicx}
\usepackage{times}
\usepackage{cite}
\usepackage{amsmath}
\usepackage{array}
\usepackage{amssymb}

\usepackage{bbm}

\usepackage{stfloats}
\usepackage{float}
\usepackage{rotating,threeparttable,booktabs}
\usepackage{bm}
\usepackage{dcolumn,booktabs}
\usepackage{multirow}
\usepackage{graphicx}
\usepackage{subfigure}
\usepackage{color}
\usepackage{epsfig}
\usepackage{epstopdf}
\usepackage{amsmath,amsthm,amssymb,amsfonts}
\makeatletter
\thm@headfont{\sc}
\makeatother

\IEEEoverridecommandlockouts

\begin{document}
\title{DC-Informative Joint Color-Frequency Modulation for Visible Light Communications}


\author{Qian Gao, \IEEEmembership{Member,~IEEE},~Rui Wang, \IEEEmembership{Member,~IEEE}, \\Zhengyuan Xu, \IEEEmembership{Senior Member,~IEEE} and Yingbo Hua, \IEEEmembership{Fellow,~IEEE}
\IEEEcompsocitemizethanks{\IEEEcompsocthanksitem Q.~Gao, Z.~Xu are with the The University of Science and Technology of China, Hefei, China (e-mail: george19870321@gmail.com; xuzy@ustc.edu.cn). \IEEEcompsocthanksitem Rui Wang is with Tongji University, Shanghai, China (e-mail: liouxingrui@gmail.com). \IEEEcompsocthanksitem Yingbo Hua is with University of California, Riverside, California, USA (e-mail: yhua@ee.ucr.edu).}}

\maketitle

\IEEEpeerreviewmaketitle

\begin{abstract}
In this paper, we consider the problem of constellation design for a visible light communication (VLC) system using red/green/blue light-emitting diodes (RGB LED), and propose a method termed DC-informative joint color-frequency modulation (DCI-JCFM). This method jointly utilizes available diversity resources including different optical wavelengths, multiple baseband subcarriers, and adaptive DC-bias. Constellation is designed in a high dimensional space, where the compact sphere packing advantage over lower dimensional counterparts is utilized. Taking into account multiple practical illumination constraints, a non-convex optimization problem is formulated, seeking the least error rate with a fixed spectral efficiency. The proposed scheme is compared with a decoupled scheme, where constellation is designed separately for each LED. Notable gains for DCI-JCFM are observed through simulations where balanced, unbalanced and very unbalanced color illuminations are considered.
\end{abstract}
\begin{IEEEkeywords}
\textbf{Constellation design, visible light communication, DC-informative, joint color-frequency, IM/DD.}
\end{IEEEkeywords}

\IEEEpeerreviewmaketitle

\section{Introduction}
To satisfy the increasingly higher data rate demands, visible light communication (VLC) has drawn tremendous interest from both industry and academia as a promising complement to traditional radio frequency communication (RFC) that suffers from spectrum saturation \cite{Komine,Raja,Elgala}. The maturing of LED manufacturing techniques during the recent decade largely boosts the trend of replacing traditional lighting systems with LED alternatives for both indoor and outdoor illumination purposes, and the resulting infrastructures are ready for deployment of VLC. It is a low-cost technology where one can use the simple intensity modulation and direct detection (IM/DD) techniques. In addition, one can enjoy a bunch of additional advantages such as eye-safety, high security and causing no electromagnetic inference.

VLC is both unique from and similar to RFC. With regard to the uniqueness of VLC, it only allows positive and real signals to drive the LEDs as intensities (a non-informative DC-bias is typically used); its channel especially for indoor environment is much more slower varying than a RFC counterpart; baseband waveforms modulates the LEDs directly instead of being up-converted first, etc. As for the similarity between VLC and RFC, many existing RFC techniques can be applied, although possibly with non-straightforward modifications to VLC systems, e.g. optical multiple input multiple output (O-MIMO) \cite{4Zeng09}, optical orthogonal frequency division multiplexing (O-OFDM) \cite{5Jean,Dimitrov,Shlomi}, and other advanced signal processing techniques \cite{You01,Teramoto05,Karout1,Monteiro,Qian,Cossu,Bai12,Vucic}. These (and relevant) works nicely take advantage of various diversities a VLC system provides, such as spatial diversity, frequency diversity, color diversity and adaptive DC-bias, to improve system performance. It is worth noting that the color diversity and adaptive DC-bias configuration are specific to VLC.

The motivation behind this work is to exploit the benefits of various diversities jointly for a very power efficient VLC, while the focus of this paper is on the problem of constellation design in high dimensional space. This space is formed by several dimensions of freedoms including adaptive DC-bias, baseband subcarriers and multiple wavelengths corresponding to R/G/B LED lights. According to the fundamental idea that spheres (i.e., constellation points) can pack more compactly in a higher dimensional  space, a constellation with larger minimum Euclidean distance (MED) can be expected in a higher dimensional space. \footnote{The system symbol error rate (SER) is governed by the MED for working electrical SNRs for VLC \cite{Qian}.}.  This MED maximization problem is formulated in a non-convex optimization form, and is then relaxed to a convex optimization problem by a linear approximation method. Key practical lighting requirements are taken into account as constraints, e.g., the optical power constraint, average color constraint, non-negative intensity constraint, color rendering index (CRI) and luminous efficacy rate (LER) requirements \cite{CIE,Stimson}.

\begin{figure*}[bp]
\centering
\includegraphics*[width=18cm]{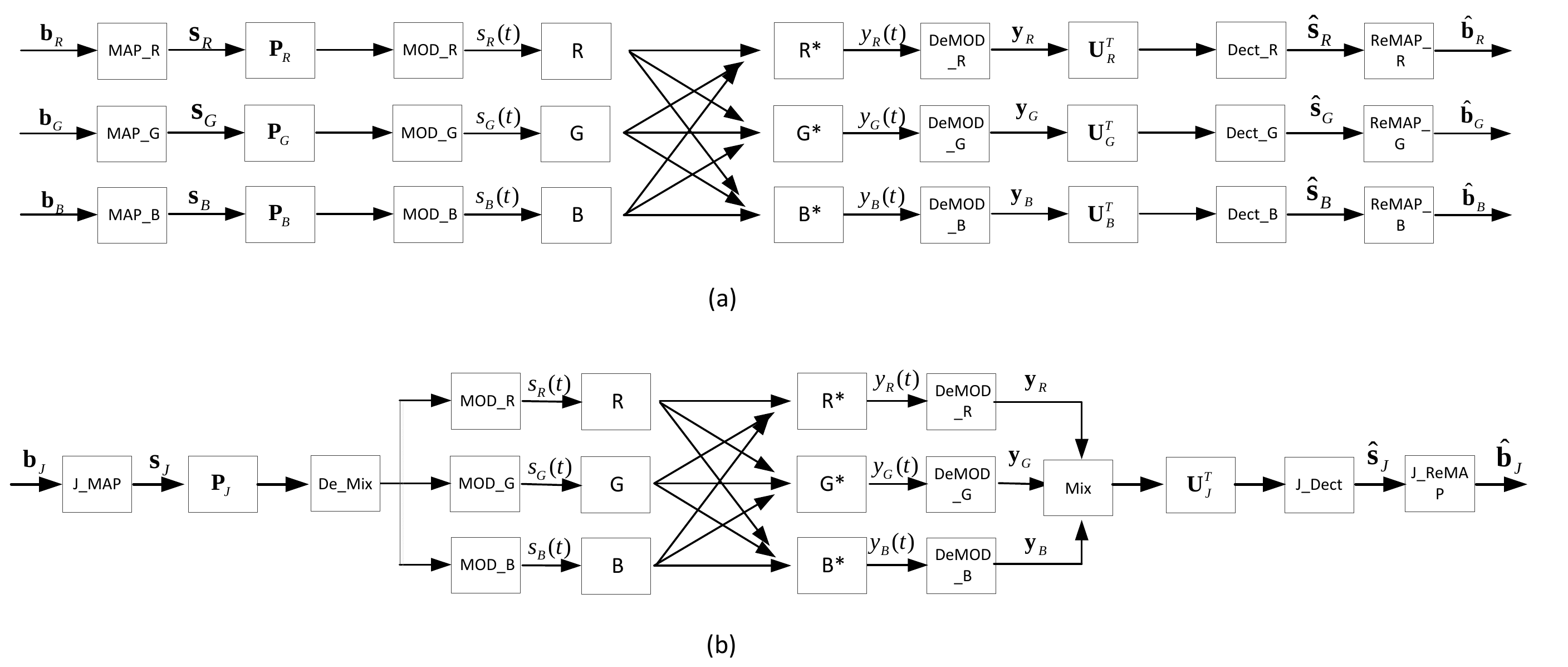}
\caption{(a). System block diagram of a decoupled system; (b). System block diagram of the DCI-JCFM. \big($(Re)MAP_{R/G/B}$: bits and constellation (Re)mapper for R/G/B tunnel; $(De)MOD_{R/G/B}$: (De)modulator for R/G/B tunnel; ${R/G/B}^*$: Photo Detector and color filter for R/G/B tunnel; $J_{(Re)MAP}$: joint (Re)mapper; $J_{Dect}$: joint symbol detector \big)}\label{fig1}
\end{figure*}

For RFC, one well-known shortcoming with utilizing multiple subcarriers is the excessive peak-to-average power ratio (PAPR) problem introduced, which can cause severe nonlinear distortion to degrade system performance. Plenty of methods have been proposed to reduce PAPR (see \cite{Han} and references therein). In fact, when using multiple subcarriers for VLC, high PAPR is also a very severe issue, due to the limited linear dynamic range of amplifiers and LEDs. This paper shows that such distortion can be avoided by formulating the dynamic range requirement as (convex) constraints of the optimization problem. In such way distortion control becomes a offline process or a by-product of constellation design.

The remainder of this paper is organized as follows.
In Section II, we first provide an overview of DC-informative modulation schemes for optical communications.  DC-informative multicarrier modulation is introduced as a power efficient alternative to traditional non-DC-informative optical OFDM configurations.
In Section III, we propose the DCI-JCFM method for systems with RGB LED. The key lighting constraints including dynamic range control are discussed. The cases of ``Balanced'', ``Unbalanced'' and ``Very Unbalanced'' systems are introduced.
In Section IV, we discuss the pros and cons of using dynamic range constraint, short time PAPR constraint and long time PAPR constraint.
In Section V, we provide simulation results to verify the significant performance gains of the proposed method over a decoupled method for balance, unbalance, and very unbalanced systems.
Finally, Section VI provides conclusions.

\vspace{.1in}

\section{An Overview of DC-informative Modulation for Optical Communications}

Optical communications based on IM/DD has a unique feature of requiring all signals modulating the LEDs to be positive (and real), so multiple schemes are proposed accordingly such as the well-know asymmetrically clipped optical OFDM (ACO-OFDM), DC-biased optical OFDM (DCO-OFDM) and optical multisubcarrier modulation (MSM). These schemes all discard the DC-bias at the receiver, which causes significant power loss. The DC-informative modulation schemes were then proposed such that $100\%$ optical power is used for data transmission (see \cite{Karout1} for single carrier case and \cite{Qian} for multiple carrier selective fading case). To be specific, consider the channel model
\begin{equation}
y(t)=\gamma\eta s_i(t)\ast h(t)+v(t)\qquad i\in[1,N_c], \label{1}
\end{equation}
where $s_i(t)$ is a symbol waveform mapped from $\mathbf{b}_i$ that contains $N_b$ bits of information, $\ast$ denotes the convolution operator, $h(t)$ is either flat-fading or selective-fading channel, $v(t)$ denotes white noise, $y(t)$ is the received signal, $\eta$ and $\gamma$ are electrical-to-optical and optical-to-electrical conversion factors respectively \footnote{We assume $\gamma\eta=1$ with out loss of generality (w.o.l.g).}, and $N_c=2^{N_b}$ stands for the constellation size. The key feature of a DC-informative modulation is that the following basis are used {\it jointly} to carry information
\begin{equation}
\phi_1(t)=\sqrt{\frac{1}{T_s}}\Pi(\frac{t}{T_s}),\label{10}
\end{equation}
\begin{equation}
\phi_{2k}(t)=\sqrt{\frac{2}{T_s}}\cos(2\pi f_kt)\Pi(\frac{t}{T_s})~~ k=1,2,\ldots,K,\label{11}
\end{equation}
\begin{equation}
\phi_{2k+1}(t)=\sqrt{\frac{2}{T_s}}\sin(2\pi f_kt)\Pi(\frac{t}{T_s})~~ k=1,2,\ldots,K,\label{12}
\end{equation}
where both I and Q channels are used. $\phi_1(t)$ is the DC-bias basis, $T_s$ is the symbol interval, $f_k=\frac{k}{T_s}$ is the $k$-th subcarrier, $K$ is the total number of subcarriers,  and a rectangular pulse-shaper is used
\begin{equation}
\Pi(t) = \left\{
\begin{array}{rl}
1, & \text{if} ~0\leq t<1\\
0, & \text{otherwise}.
\end{array} \right.\label{13}
\end{equation}
The relationship between symbol waveforms $s_i(t)$ and constellation points $\mathbf{s}_i=[s_{1,i},s_{2,i},\ldots,s_{2K+1,i}]$ is
\begin{equation}
s_i(t)=\underbrace{s_{1,i}\phi_1(t)}_{\text{Adaptive Bias}}+s_{2,i}\phi_2(t)+\ldots+s_{2K+1,i}\phi_{2K+1}(t).\label{6}
\end{equation}
And one of the reasonable goals is to minimize the SER subject to fixed electrical/optical power by properly design the constellation matrix
\begin{equation*}
\mathcal{S}=
\begin{bmatrix}
s_{1,1} & s_{1,2} & \ldots & s_{1,N_c}\\
s_{2,1} & s_{2,2} & \ldots & s_{2,N_c}\\
\vdots & \vdots & \ddots & \vdots\\
s_{2K+1,1} & s_{2K+1,2} & \ldots & s_{2K+1,N_c}\\
\end{bmatrix},
\end{equation*}
where each column of $\mathcal{S}$ is a constellation point, and the MED of all columns should be maximized to reach the goal. We typically stack the columns into a single vector instead, i.e.
\begin{equation}
\mathbf{s}_J=[\mathbf{s}_1^T~\mathbf{s}_2^T~\ldots~\mathbf{s}_{N_c}^T]^T,
\end{equation}
for simplification of the formulation of the optimization problem discussed in Section II.

\section{DC-informative Joint Color-Frequency Modulation with RGB LEDs}
Based on the idea discussed in Section II, we propose two constellation design methods taking advantage of the informative DC-bias for a visible light communication system employing one RGB LED as shown by Fig.1. If the inputs into R/G/B carry independent bit information as shown by Fig.1(a), it is termed a decoupled scheme. In comparison, if the inputs into R/G/B modulators only carry information jointly, it is a joint scheme instead as shown by Fig.1(b). In other words, although for a joint scheme still three modulators are used to create continuous domain waveforms to drive corresponding LEDs, information cannot be estimated though recovery of a single (or any pair) of them.

It is observed that the joint scheme may utilize four types of diversities during per channel use, including frequency diversity, color(wavelength) diversity, adaptive DC, and spatial diversity. Spatial diversity can be achieved by extending from employing only one RGB LED to include $N$ ones, which is out of scope of this paper. We only emphasize the first three diversities here. This scheme is termed DC-informative joint color-frequency modulation (DCI-JCFM).

The decoupled system shown in Fig.1(a) works as follows: At the transmitter-side three independent bit stream $\mathbf{b}_{x,i},~ x\in[\text{red,green,blue}]~i\in[1,N_c]$, of length $N_x$ are mapped to corresponding constellation points $\mathbf{s}_{x,i}$ of size $(2K+1)\times 1$ first, which are modulated separately to generate continuous symbol waveform (current) $s_{x,i}(t)$ by \eqref{6} for each tunnel. If cross-talks exist for any tunnel, a corresponding precoder $\mathbf{P}_x$ needs to be applied before modulation. Waveforms $s_{x,i}(t)$ are then electrical-to-optical converted to intensity signals $\eta s_{x,i}(t)$ to drive the LEDs. At the receiver-side, photo detectors of each tunnel collect the waveforms (convoluted with channel and corrupted by noise). The received signal is sent through red, green, and blue color filters respectively and after optical-to-electrical conversion waveforms $y_{x,i}(t)$ are obtained. Then $2K+1$ matched filters are employed for each tunnel to demodulate $y_{x,i}(t)$ to obtain signal vector $\mathbf{y}_{x,i}$. Three symbol detectors
follow to provide $\hat{\mathbf{s}}_{x,i}$, estimates of the symbol vectors, which are de-mapped separately and
the estimate of original bit sequences $\mathbf{b}_{x,i}$ are finally obtained. If cross-talks exist, post-equalizers $\mathbf{U}_x^T$ are applied before the symbol detectors.

System using DCI-JCFM as shown by Fig.1(b) works differently. A joint bit sequence \begin{equation}
\mathbf{b}_{J,i}=[\mathbf{b}_{R,i}^T~\mathbf{b}
_{G,i}^T~\mathbf{b}_{B,i}^T]^T,
\end{equation}
is firstly mapped jointly to a constellation point $\mathbf{s}_{J,i}$ of size $(6K+3)\times 1$.
Then $\mathbf{s}_{J,i}$ is converted by a joint modulator to the continuous domain to obtain $s_{J,i}$ through
\begin{align}
s_{J,i}(t)&=\underbrace{\sum_{p=1}^3 s_{(p-1)(2K+1)+1,i}\phi_1(t)}_{\text{Adaptive R/G/B Bias}}\notag\\
&+\sum_{p=1}^3\sum_{k=2}^{2K+1}s_{(p-1)(2K+1)+k,i}\phi_k(t).
\end{align}
If cross-talks exist, a joint precoder $\mathbf{P}_J$ is applied before modulation. Also we observe the expectation of $s_{J,i}(t)$ as follows
\begin{align}
\mathbb{E}[s_{J,i}(t)]&=\sum_{p=1}^3 s_{(p-1)(2K+1)+1,i}\phi_1(t),
\end{align}
since all non-DC basis has zero time averages. Therefore, both the average optical power and average color of system
are determined solely by these three adaptive DC-bias. While the dynamic range of waveform, instead, is influenced by
all subcarriers of all LEDs.

For our design, we will demonstrate with a line-of-sight (LOS) scenario when channel has cross-talks, due to the imperfectness
of receiver color filters. The discrete channel model can be written as follows \cite{Monteiro}
\vspace{-0.02in}
\begin{align}
\mathbf{y}&=\mathbf{H}\mathbf{s}+\mathbf{n}\notag\\
&=\begin{bmatrix}
\mathbf{y}_{R} \\
\mathbf{y}_{G} \\
\mathbf{y}_{B}
\end{bmatrix}
=\begin{bmatrix}
\mathbf{I} & \mathbf{O} & \mathbf{O}\\
\mathbf{O} & (1-2\epsilon)\mathbf{I} & \epsilon\mathbf{I}  \\
\mathbf{O} & \epsilon\mathbf{I} & (1-2\epsilon)\mathbf{I}
\end{bmatrix}
\begin{bmatrix}
\mathbf{s}_R \\
\mathbf{s}_G \\
\mathbf{s}_B
\end{bmatrix}+
\begin{bmatrix}
\mathbf{n}_R \\
\mathbf{n}_G \\
\mathbf{n}_B
\end{bmatrix},\label{2}
\end{align}
where $\epsilon\in[0,0.5]$ is termed the ``cross-talk index (CI)'' and $\mathbf{n} \sim {\cal N}({\bf 0}, {\bf I} \cdot N_0)$.

\subsection{The objective function}
With working SNRs for VLC which are typically medium-to-high, the minimum Euclidean distance between constellation pairs governs SER. Therefore, we seek to minimize the system SER by maximizing the minimum Euclidean distance, through carefully design the constellation vector $\mathbf{s}_J$ subject to key lighting constraints. For a constellation containing $N_c$ points, distances of a total of $N_c(N_c-1)/2$ pairs have to be constrained as follows \cite{Beko12}
\begin{equation}
\mathbf{s}_J^T\mathbf{F}_l\mathbf{s}_J\geq d^2_{min},\label{8}
\end{equation}
where we define
\begin{equation}
\mathbf{F}_{l(p,q)}=\mathbf{E}_{pq},
\end{equation}
and
\begin{equation}
\mathbf{E}_p=\mathbf{e}_p^T\otimes \mathbf{I}_{N_c},
\end{equation}
where $\otimes$ denotes Kronecker product, $\mathbf{e}_p$ is the $p$-th column of identity matrix $\mathbf{I}_{6K+3}$, and
\begin{equation}
\mathbf{E}_{pq}=\mathbf{E}_p^T\mathbf{E}_p-\mathbf{E}_p^T\mathbf{E}_q-
\mathbf{E}_q^T\mathbf{E}_p+\mathbf{E}_q^T\mathbf{E}_q,
\end{equation}
where $l\cong(p-1)N_c-\frac{p(p+1)}{2}+q,~p,q\in1,2,\ldots,N_c,~p<q$.

The distance constraints are nonconvex in $\mathbf{s}_J$. We choose to use the follows linear approximation at point $\mathbf{s}_J^{(0)}$
\begin{align}
\mathbf{s}_J^T\mathbf{F}_l\mathbf{s}_J\cong 2\mathbf{s}_J^{(0)T}\mathbf{F}_l\mathbf{s}_J-
\mathbf{s}_J^{(0)T}\mathbf{F}_l\mathbf{s}_J^{(0)}\geq d^2_{min},\qquad \forall l.\label{28}
\end{align}
\subsection{Practical lighting requirements}
For our design problem, practical lighting issue considered include average optical power, average illumination color, LER and CRI, non-negative intensity, and flickering-free requirements. The first three requirements can be constraint using only one equation written as follow
\begin{equation}
P_o\cdot\mathbf{s}_{avg}=\frac{1}{N_c}\mathbf{J}\mathbf{s}_J,
\end{equation}
where $P_o$ is the average optical power of a RGB LED, $\mathbf{J}$ is a selection matrix (containing only ones and zeros) adding up R/G/B components in $\mathbf{s}_J$ respectively by a multiplication of each row with it, $\mathbf{s}_{avg}=[s_R~s_G~s_B]^T$ is termed the average color ratio vector and the follow equation holds
\begin{equation}
s_R+s_G+s_B=1.
\end{equation}
Thus the optical power and illumination color requirements are constrained together. The luminous efficacy rate and color rendering index requirements can be satisfied by properly choosing $\mathbf{s}_{avg}$.

\subsection{Dynamic range requirement}
Since the linear dynamic range of LEDs are limited, the ranges of signal for each LED have to be constrained to avoid nonlinear distortion, i.e.
\begin{equation}
0\leq s_{x,i}(t)\leq I_U,\qquad \forall x,i,\label{13}
\end{equation}
where $I_U$ is the highest current level and for simplicity we have assumed that red, green and blue LEDs have the same dynamic range. We propose to constrain dynamic range of a sequence of sampled signal
\begin{equation}
0\leq s_{x,i}(t_n)\leq I_U,\qquad \forall x,i,\label{14}
\end{equation}
and $t_n$ is picked as
\begin{equation}
t_n=\frac{nT_s}{2KN_o}, \qquad n=0,1,\ldots,N,
\end{equation}
where $N_o$ is the oversampling rate, $N=2KN_o$, and $N+1$ is the total number of sample points. It should be noted that although \eqref{14} does not guarantee \eqref{13}, which means the continuous signal waveforms designed subject to \eqref{14} could result in negative amplitudes in between the sample instances, the negative peak is very small compare to the dynamic range of signal. We can compensate this effect by adding a small post DC-bias after obtaining an optimized constellation.

Therefore, we can formulate this point-wise dynamic range constraints as follows

\begin{equation}
\mathbf{u}_n^T\mathbf{K}_x\mathbf{J}_i\mathbf{s}_J\geq 0, \qquad \forall (x,i,n)
\end{equation}
\begin{equation}
\mathbf{u}_n^T\mathbf{K}_x\mathbf{J}_i\mathbf{s}_J\leq I_u, \qquad \forall (x,i,n)
\end{equation}
where $\mathbf{J}_i$ selects the $i$-th constellation point and $\mathbf{K}_x$ selects the corresponding coefficients for color $x$, $\mathbf{u}_n=[u_{n,0},u^c_{n,1},u^s_{n,1},\ldots,u^c_{n,K},u^s_{n,K}]^T$, $u_{n,0}=\sqrt{1/T_s}$, $u^c_{n,k}=\sqrt{2/T_s}\cos(2\pi f_kt_n)$, and $u^s_{n,k}=\sqrt{2/T_s}\sin(2\pi f_kt_n)$.

\subsection{Problem formulation}
We first formulate the optimization problem when there is no cross-talks among different colored LEDs, i.e. $\mathbf{H}=\mathbf{I}$, as follows

\begin{equation}
\begin{aligned}
& \underset{\mathbf{s}_J,d_{min,J}}{\text{maximize}}
& & d_{min,J} \\
& \text{s.t.}
& & P_o\cdot\mathbf{s}_{avg}=\frac{1}{N_c}\mathbf{J}\mathbf{s}_J\\
&&& 2\mathbf{s}_J^{(0)T}\mathbf{F}_l\mathbf{s}_J-
\mathbf{s}_J^{(0)T}\mathbf{F}_l\mathbf{s}_J^{(0)}\geq d^2_{min,J}\qquad \forall l.\\
&&& \mathbf{u}_n^T\mathbf{K}_x\mathbf{J}_i\mathbf{s}_J\geq 0 \qquad \forall (x,i,n)\\
&&& \mathbf{u}_n^T\mathbf{K}_x\mathbf{J}_i\mathbf{s}_J\leq I_u \qquad \forall (x,i,n),
\end{aligned}
\end{equation}
which is convex in $\mathbf{s}_J$ and $d_{min}$, and specialized solver such as CVX toolbox for MATLAB can be utilized \cite{cvx}. Start from initial point $\mathbf{s}_J^{(0)}$, the scheme can iteratively converge to a local optima with each run. The best constellation is chosen from local optimal constellation obtained from multiple runs.

When the channel suffers from cross-talks, we choose to deal with it by employing the well-known singular value decomposition (SVD) based pre-equalizer $\mathbf{P}=\mathbf{V}\mathbf{S}^{-1}$ and post-equalizer $\mathbf{U}^H$ for our system, where $\mathbf{H}=\mathbf{U}\mathbf{S}\mathbf{V}^H$. Constellation is designed by an optimization with transformed constraints as follow

\begin{equation}
\begin{aligned}
& \underset{\mathbf{s}_J,d_{min,J}}{\text{maximize}}
& & d_{min,J} \\
& \text{s.t.}
& & P_o\cdot\mathbf{s}_{avg}=\frac{1}{N_c}\mathbf{J}\mathbf{P}_J\mathbf{s}_J\\
&&& 2\mathbf{s}_J^{(0)T}\mathbf{F}_l\mathbf{s}_J-
\mathbf{s}_J^{(0)T}\mathbf{F}_l\mathbf{s}_J^{(0)}\geq d^2_{min,J}\qquad \forall l.\\
&&& \mathbf{u}_n^T\mathbf{K}_x\mathbf{J}_i\mathbf{P}_J\mathbf{s}_J\geq 0 \qquad \forall (x,i,n)\\
&&& \mathbf{u}_n^T\mathbf{K}_x\mathbf{J}_i\mathbf{P}_J\mathbf{s}_J\leq I_u \qquad \forall (x,i,n),
\end{aligned}\label{20}
\end{equation}
where $\mathbf{P}_J=\mathbf{I}_{N_c}\otimes\mathbf{P}$ is defined \footnote{$\otimes$ is the Kronecker product.} and apparently this optimization is convex as well.

For the decoupled system, three independent problems can be formulated to find the MEDs for each color, i.e.

\begin{equation}
\begin{aligned}
& \underset{\mathbf{s}_{R/G/B},d_{min,R/G/B}}{\text{maximize}}
& & d_{min,R/G/B} \\
& \text{s.t.}
& & P_o\cdot s_{R/G/B}=\frac{1}{N_c}\mathbf{j}^T\mathbf{P}_{R/G/B}\mathbf{s}_{R/G/B}\\
&&& 2\mathbf{s}_{R/G/B}^{(0)T}\tilde{\mathbf{F}}_l\mathbf{s}_{R/G/B}-
\mathbf{s}_{R/G/B}^{(0)T}\tilde{\mathbf{F}}_l\mathbf{s}_{R/G/B}^{(0)}\\
&&&\geq d^2_{min,R/G/B}\qquad \forall l.\\
&&& \mathbf{u}_n^T\tilde{\mathbf{J}}_i\mathbf{P}_{R/G/B}\mathbf{s}_{R/G/B}\geq 0 \qquad \forall (i,n)\\
&&& \mathbf{u}_n^T\tilde{\mathbf{J}}_i\mathbf{P}_{R/G/B}\mathbf{s}_{R/G/B}\leq I_{u,R/G/B} \qquad \forall (i,n),
\end{aligned}
\end{equation}
where $d_{min,R/G/B}$, $\mathbf{j}$, $\mathbf{P}_{R/G/B}$, $\tilde{\mathbf{F}}_l$, $\tilde{\mathbf{J}}_i$, and $I_{u,R/G/B}$ are defined in a similar manner with corresponding parameters in \eqref{20}. For brevity we omit the explicit definitions.

\section{Dynamic Range VS PAPR Constraints}
In fact, although a hard constraint on the dynamic range of symbol waveforms can help avoid non-linear distortion completely, it may bring with side effects such as excessive power efficiency decrease. This is particularly true if only one or few symbol waveforms have notably larger dynamic range than the majority. In such case, one can consider using certain PAPR constraint to replace the dynamic range constraint. In other words, there is a tradeoff between allowable PAPR and power efficiency.

Two types of PAPR constraints (for each LED light) can be considered. One is the so-called long-term PAPR (L-PAPR), i.e. the ratio of the peak power of all waveforms and the time average of them. Assuming no cross-talks, the L-PAPR constraint for LED $x$ can be written as
\begin{align}
\Phi_x(\mathbf{s}_J)&=\frac{[\max_{i,n}(\mathbf{u}_n^T\mathbf{K}_x\mathbf{J}_i\mathbf{s}_J)]^2}
{\mathbf{s}_J^T\mathbf{s}_J/N_c}\leq \beta_x\notag\\
&=\frac{N_c[\max_{i,n}(\mathbf{u}_n^T\mathbf{K}_x\mathbf{J}_i\mathbf{s}_J)]^2}
{\mathbf{s}_J^T\mathbf{s}_J}\leq\beta_x,\label{27}
\end{align}
where $\beta_x$ is the required L-PAPR for LED $x$. Thus, a set of constraints can be formulated as follows
\begin{equation}
\mathbf{u}_n^T\mathbf{K}_x\mathbf{J}_i\mathbf{s}_J-\sqrt{\frac{\beta_x\mathbf{s}_J^T\mathbf{s}_J}{N_c}}\leq 0\qquad \forall (i,n),
\end{equation}
which is non-convex in $\mathbf{s}_J$. A way to deal with this is to use a similar linear approximation as in \eqref{28} at the same initial point $\mathbf{s}_J^{(0)}$. The above constraints are thus transformed as follows
\begin{equation}
\mathbf{u}_n^T\mathbf{K}_x\mathbf{J}_i\mathbf{s}_J-
\sqrt{\frac{\beta_x}{N_c}}(\mathbf{s}_J^{(0)T}\mathbf{s}_J^{(0)})^{-\frac{1}{2}}
\mathbf{s}_J^{(0)T}
(\mathbf{s}_J-\mathbf{s}_J^{(0)})\leq 0~\forall (i,n).
\end{equation}

The other is the individual PAPR (I-PAPR), i.e. the ratio of the peak power of each waveform and the average power of it. The I-PAPR for $i$-th waveform for LED $x$ can be written as
\begin{align}
\Phi_{x,i}(\mathbf{s}_J)&=\frac{[\max_{n}(\mathbf{u}_n^T\mathbf{K}_x\mathbf{J}_i\mathbf{s}_J)]^2}
{\mathbf{s}_J^T\mathbf{J}_i^T\mathbf{J}_i\mathbf{s}_J}\leq \beta_{x,i}
\end{align}

The corresponding constraint can be written as
\begin{equation}
\mathbf{u}_n^T\mathbf{K}_x\mathbf{J}_i\mathbf{s}_J-\sqrt{\beta_{x,i}\mathbf{s}_J^T\mathbf{J}_i^T\mathbf{J}_i\mathbf{s}_J}\leq 0\qquad \forall n,
\end{equation}
and a similar linear approximation process is applied to convert them to convex constraints. To the best of our knowledge, there is no comprehensive comparison on performance of systems applying those three constraints available so far.

An interesting observation recently in \cite{Li} shows that the non-linearity mitigation for an IM/DD VLC is a more involved problem than expected. The reason is that the low part of the baseband frequencies are causing larger non-linear distortion than the higher part. This effect, if taken into account along with the three constraints discussed above, is expected to make the design problem even more worthwhile to look into.

\section{Performance Evaluation}
In this section, we compare the performances of the decoupled scheme and DCI-JCFM by assessing the maximum MED and bit error rate (BER) under different channel cross-talk and color illumination assumptions \footnote{A binary switching (BSA) algorithm is applied for optimally map bit sequences to constellation points\cite{Schreckenbach}.}. Each constellation point is assumed to have equal probability of transmission, and the union bound for SER of both the scheme can be written as \cite[Eq.25]{Karout1}
\begin{equation}
P_{e,s}\approx \frac{2N_n}{N_c}Q\bigg(\sqrt{\frac{d_{min,z}^2}{2N_0}}\bigg),
\end{equation}
where $N_n$ is the number of neighbor constellation pairs \cite{Karout1} and \begin{equation}
Q(x)=\frac{1}{\sqrt{2\pi}}\int_{x}^{\infty}\exp(-t^2/2)dt
\end{equation}
denotes the Gaussian Q-function, $d_{min,z\in[J,R,G,B]}$ stands for the MED of the DCI-JCFM and MEDs for decoupled schemes. The bit error rate is thus calculated as \begin{equation}
P_{e,b}= \frac{\lambda}{N_b}P_{e,s},
\end{equation}
where $\lambda$ is the number of wrongly detected bits in each bit sequence, which can be minimized by employing the BSA mapper.

\subsection{System comparison with no channel cross-talk}

We first compare the DCI-JCFM and the decoupled scheme when channel cross-talks do not exist. To guarantee a fair comparison, the following system parameters are chosen: the number of Monte-Carlo runs for each scheme $N_M=20$, the length of bit sequence for each channel use is $N_b=6$ for DCI-JCFM and $N_{b_{R/G/B}}=2$ respectively for each tunnel of the decoupled system, the number of subcarriers for each LED is $K=2~\text{or}~3$, the average optical power $P_o=20$, the symbol interval $T_s=1$ is used\footnote{With out loss of generality $T_s=1$ is chosen, since the design is rate independent.}, the upper bound of waveform amplitude $I_U=80$, the average color ratio vector for a balanced system
\begin{equation}
\mathbf{s}_{avg,B}=[1/3,1/3,1/3]^T,
\end{equation}
for an unbalanced system
\begin{equation}
\mathbf{s}_{avg,U}=[4/9,3/9,2/9]^T,
\end{equation}
and for a very unbalanced system
\begin{equation}
\mathbf{s}_{avg,VU}=[0.7,0.15,0.15]^T.
\end{equation}

 The MEDs of the two schemes for three systems obtained through picking the best constellatio from the $20$ local optimums are summarized by Table I.

From Table I, key observations include:
a. With the DCI-JCFM, the ``joint MED'' is much larger than the R/G/B ``decoupled MEDs'', except for the cases with very unbalanced illumination. While for the very unbalanced case the blue and green tunnels could suffer from severe performance loss with the small MEDs, and therefore the DCI-JCFM is still expected to work better.
b. Larger MEDs are obtained with an increased number of subcarriers. We are only listing the cases when $K=2$ and $K=3$ for brevity, while we have observed through additional simulations that this gain continue to grow with $K$.
c. The more balanced a system is, the better performance is expected.
\begin{table}[h]
  \centering
  \caption{MED comparison with no cross-talk, DCI-JCFM (row 1$\&$2) VS Decoupled (row 3$\&$4).}
  \vspace{-0.2in}
  \begin{tabular}{cccccccc} \\ \hline
    $d_{min,z}$ & \text{Balanced} & \text{Unbalanced} & \text{Very Unbalanced} \\ \hline
    $\text{K=2}$ & 19.995 & 19.588 & 18.020 \\
    $\text{K=3}$ & 24.818 & 24.559 & 22.752  \\
    $\text{K=2}$ & [13.18,13.18,13.18] & [15.90,11.93,7.95] & [27.68,5.93,5.93] \\
    $\text{K=3}$ & [15.03,15.03,15.03] & [18.04,13.53,9.02] & [31.56,6.76,6.76] \\ \hline
  \end{tabular}
\end{table}

\subsection{DCI-JCFM performance with channel cross-talk}
With $K=2$ and other parameters given the same values as in the previous section for DCI-JCFM, we simulate to obtain the best MEDs subject to different cross-talk levels, controlled by CI varying from [0,0.2] (since with only average quality color filters CI beyond 0.2 can be avoided).

\begin{table}[h]
  \centering
  \caption{MED with different cross-talk levels, DCI-JCFM.}
  \vspace{-0.2in}
  \begin{tabular}{cccccccc} \\ \hline
    $d_{min,z}$ & \text{Balanced} & \text{Unbalanced} & \text{Very Unbalanced} \\ \hline
    $\epsilon=0$ & 19.995 & 19.588 & 18.020 \\
    $\epsilon=0.05$ & 18.834 & 18.798 & 17.014  \\
    $\epsilon=0.1$ & 16.898 & 16.835 & 16.346 \\
    $\epsilon=0.15$ & 15.334 & 15.037 & 15.192  \\
    $\epsilon=0.2$ & 14.444 &  14.250 & 14.273  \\ \hline
  \end{tabular}
\end{table}

From Table II, key observations include:
a. With increased channel cross-talks, the performances of all system degrade monotonously.
b. The performance of the balanced system remains the best with any level of channel cross-talks.
c. The DCI-JCFM is kind of robust with cross-talks, since with a severe cross-talk level, i.e. $\epsilon=0.2$, the system performance is still comparable or even better than a decoupled counterpart.

\subsection{The designed symbol waveforms with DCI-JCFM}
We pick the optimized constellation $\mathbf{s}_J^*$ designed for unbalanced system as an example in this section. If cross-talks do not exist, the corresponding subcarrier symbol waveforms $s_{R,i}(t),~s_{G,i}(t),~s_{B,i}(t)~\forall i$ obtained with DCI-JCFM are plotted in Fig. 2 - Fig. 4. With fixed optical power $P_0=20$ and varying noise power $N_0$, Fig. 5 includes bit error rate curves of the two scheme with different color illumination across selected working electrical SNRs, which is defined as
\begin{align}
\text{SNR}&=10\log_{10}\frac{\mathbb{E}(\mathbf{s}_i^T\mathbf{s}_i)}{N_0}\notag\\
&=10\log_{10}\frac{\mathbf{s}_J^T\mathbf{s}_J}{N_cN_0},
\end{align}
for the DCI-JCFM and SNR for the decoupled scheme is defined similarly.
Significant power gains of the DCI-JCFM over the decoupled scheme are observed. Also, the more unbalanced the system is, the worse performance is expected. If cross-talks exist and $\epsilon=0.1$, the corresponding subcarrier symbol waveforms are plotted in Fig. 6 - Fig. 8. With each figure, the symbol waveforms are differentiated by color. In practice, the sampled version of these waveforms can be pre-stored in the memory of a high speed waveform generator.

\begin{figure}[H]
\centerline{\includegraphics[width=1.1\columnwidth]{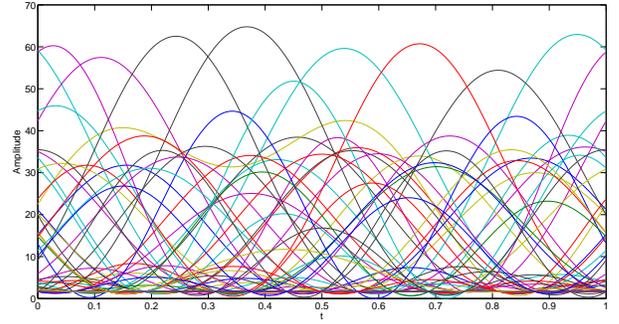}}
\vspace{-0.1in}
\caption{64 red sub waveforms for DCI-JCFM with $K=2$, no cross-talks.}
\end{figure}
\vspace{-.25in}
\begin{figure}[H]
\centerline{\includegraphics[width=1.1\columnwidth]{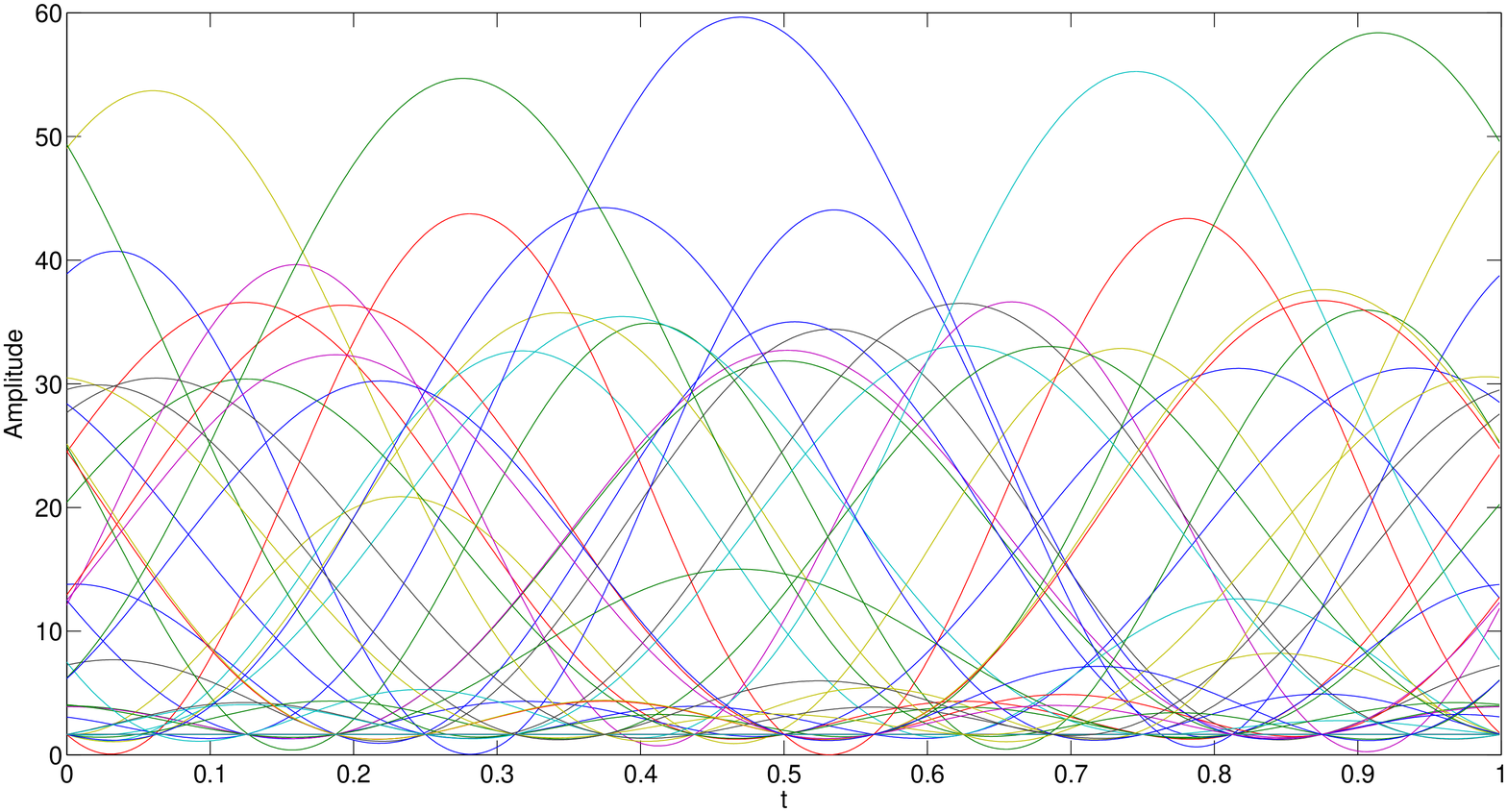}}
\vspace{-0.1in}
\caption{64 green sub waveforms for DCI-JCFM with $K=2$, no cross-talks.}
\end{figure}
\vspace{-0.05in}
\begin{figure}[H]
\centerline{\includegraphics[width=1.1\columnwidth]{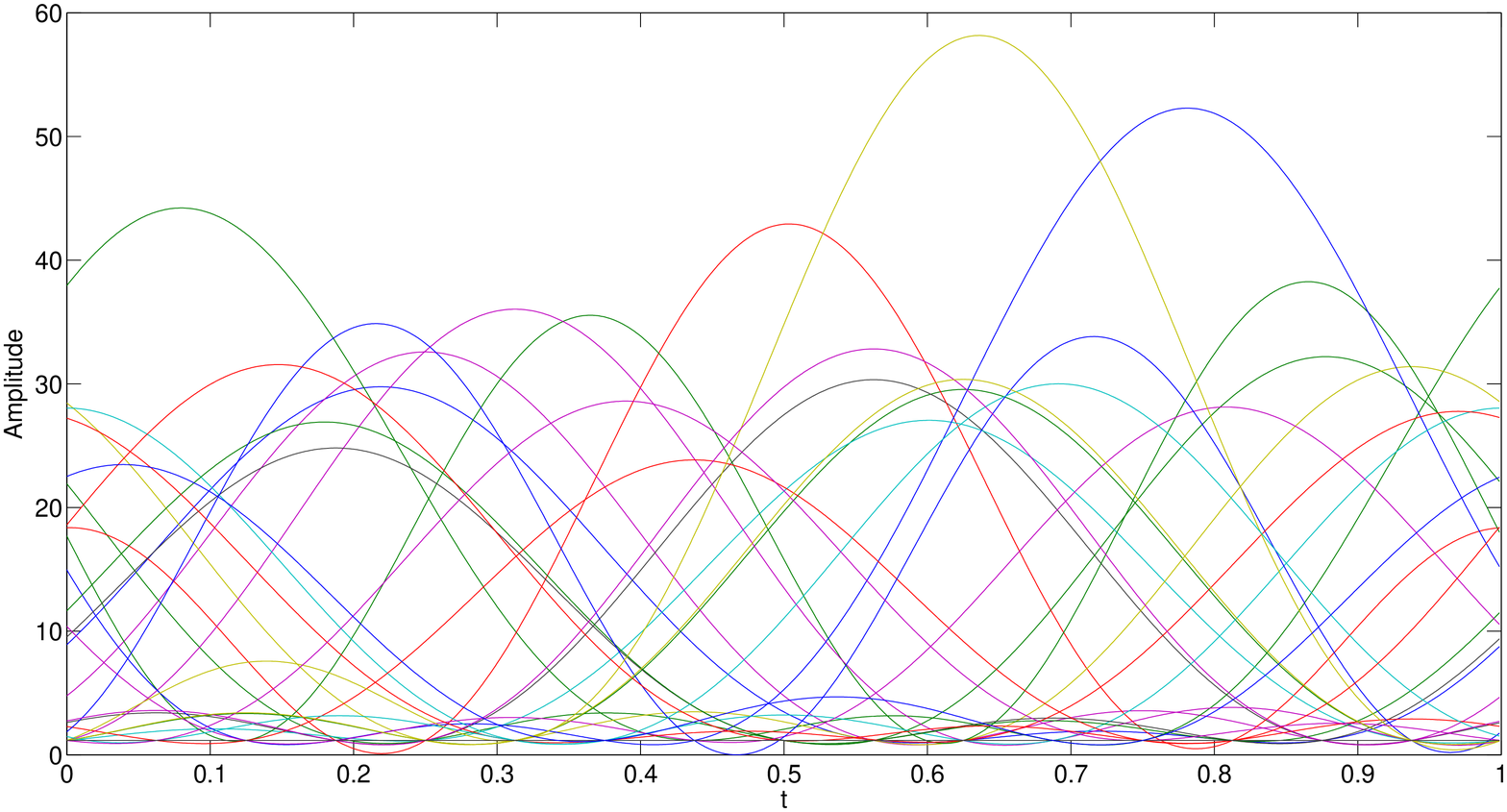}}
\vspace{-0.1in}
\caption{64 blue sub waveforms for DCI-JCFM with $K=2$, no cross-talks.}
\end{figure}
\begin{figure}[H]
\centerline{\includegraphics[width=1.1\columnwidth]{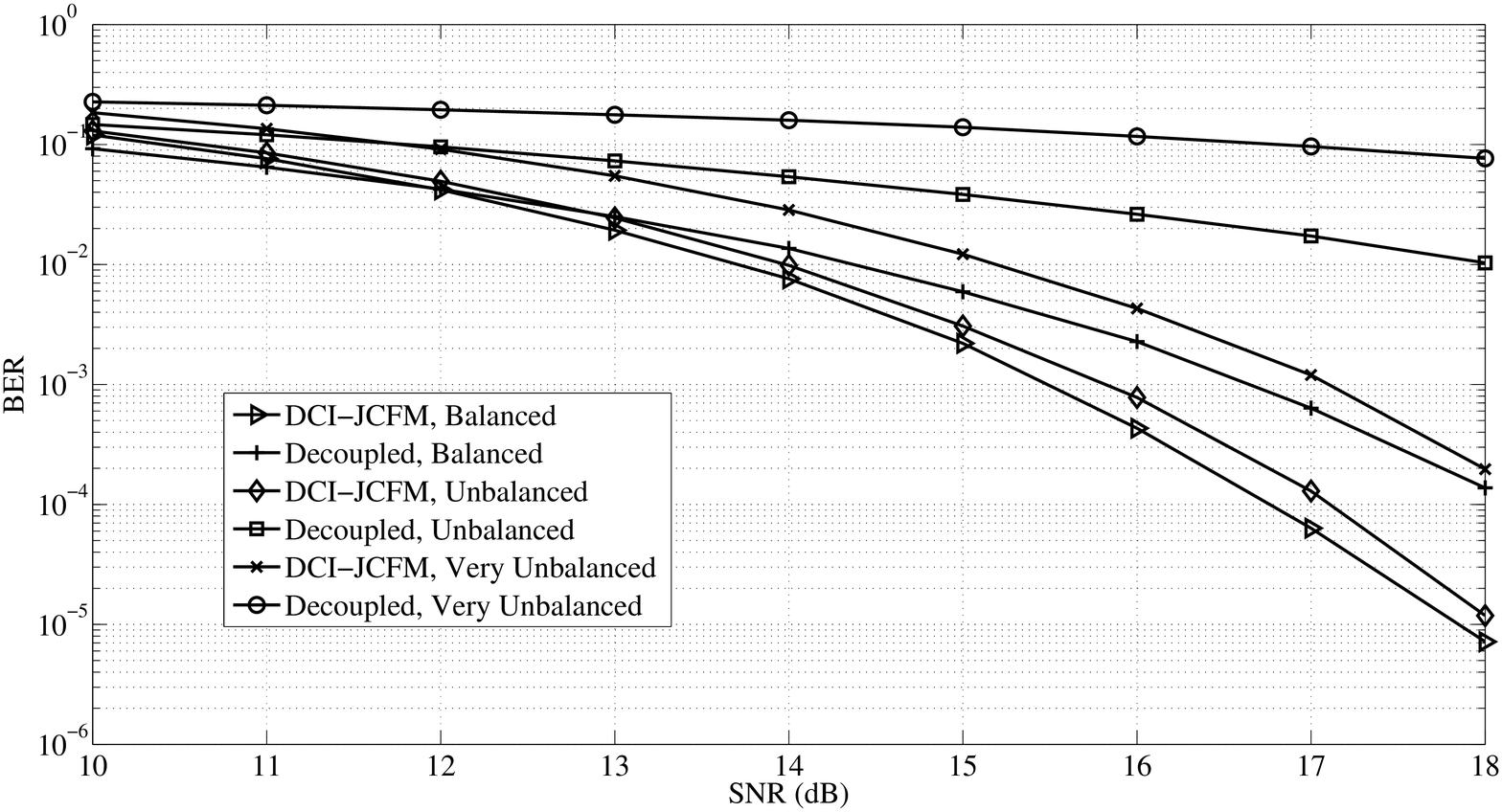}}
\vspace{-0.1in}
\caption{BER performance of DCI-JDCM and the decouple scheme.}
\end{figure}
\begin{figure}[H]
\centerline{\includegraphics[width=1.1\columnwidth]{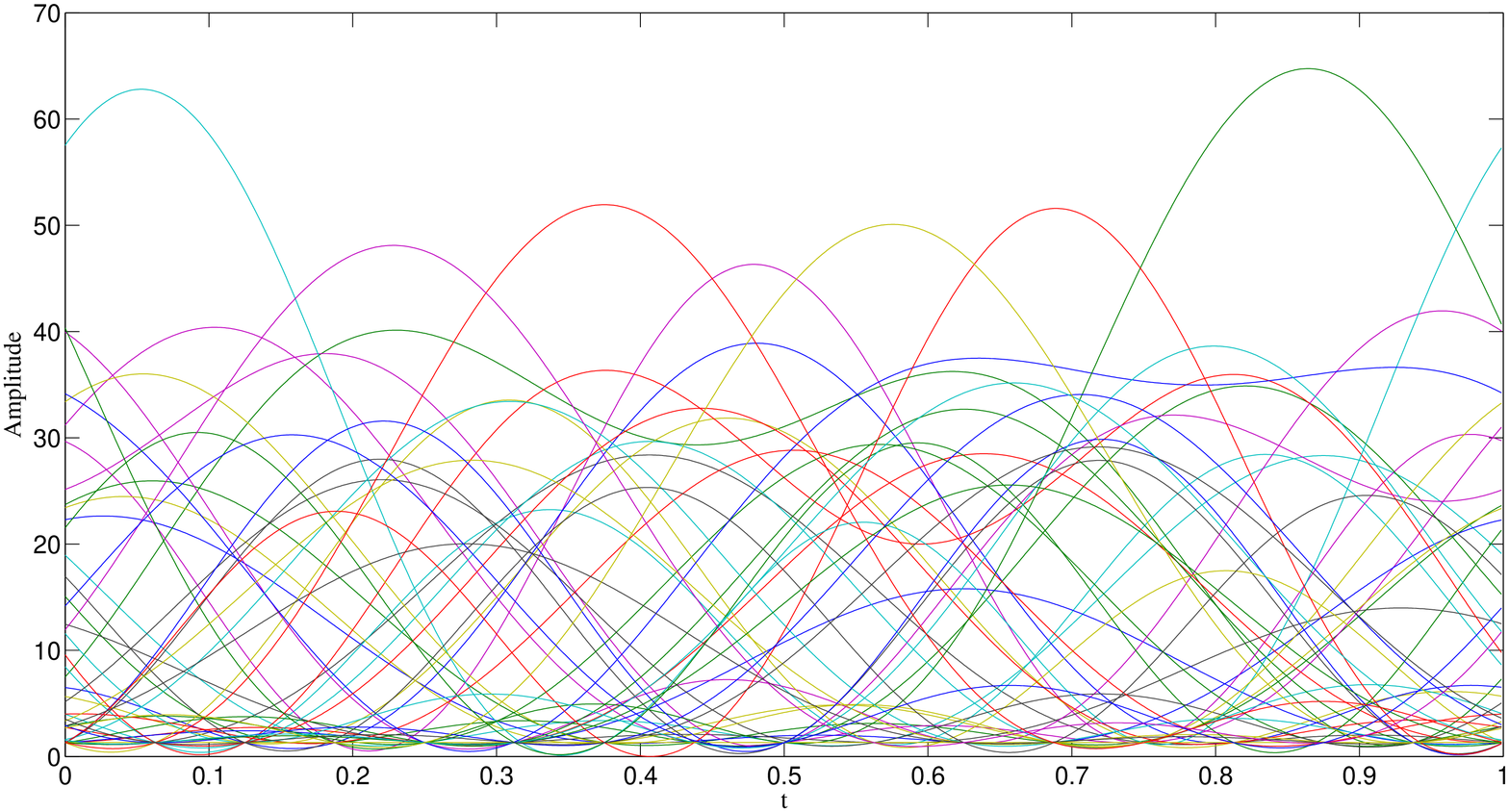}}
\vspace{-0.1in}
\caption{64 red sub waveforms for DCI-JCFM with $K=2$, $\epsilon=0.1$.}
\end{figure}
\begin{figure}[H]
\centerline{\includegraphics[width=1.1\columnwidth]{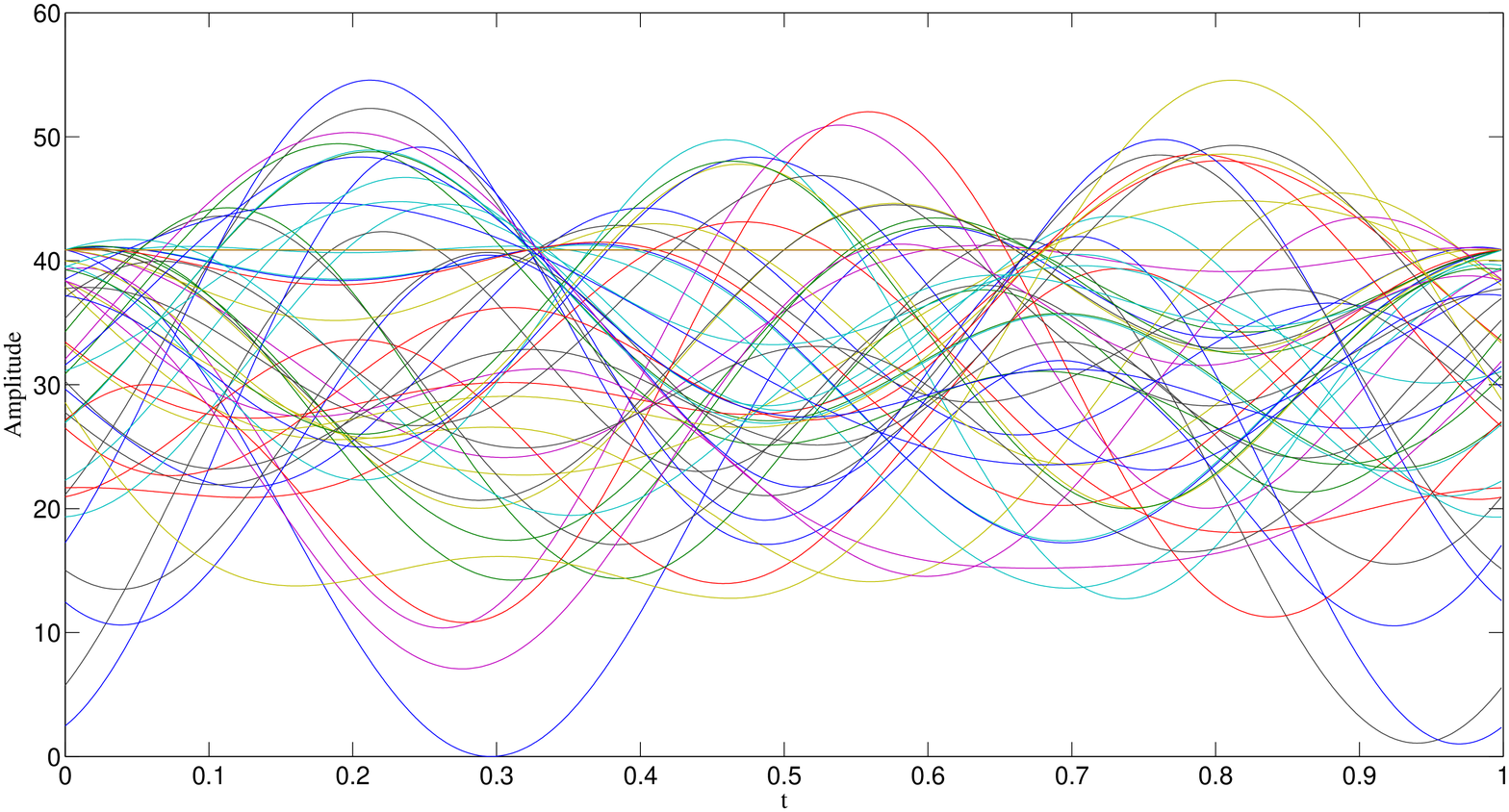}}
\vspace{-0.1in}
\caption{64 green sub waveforms for DCI-JCFM with $K=2$, $\epsilon=0.1$.}
\end{figure}
\begin{figure}[H]
\centerline{\includegraphics[width=1.1\columnwidth]{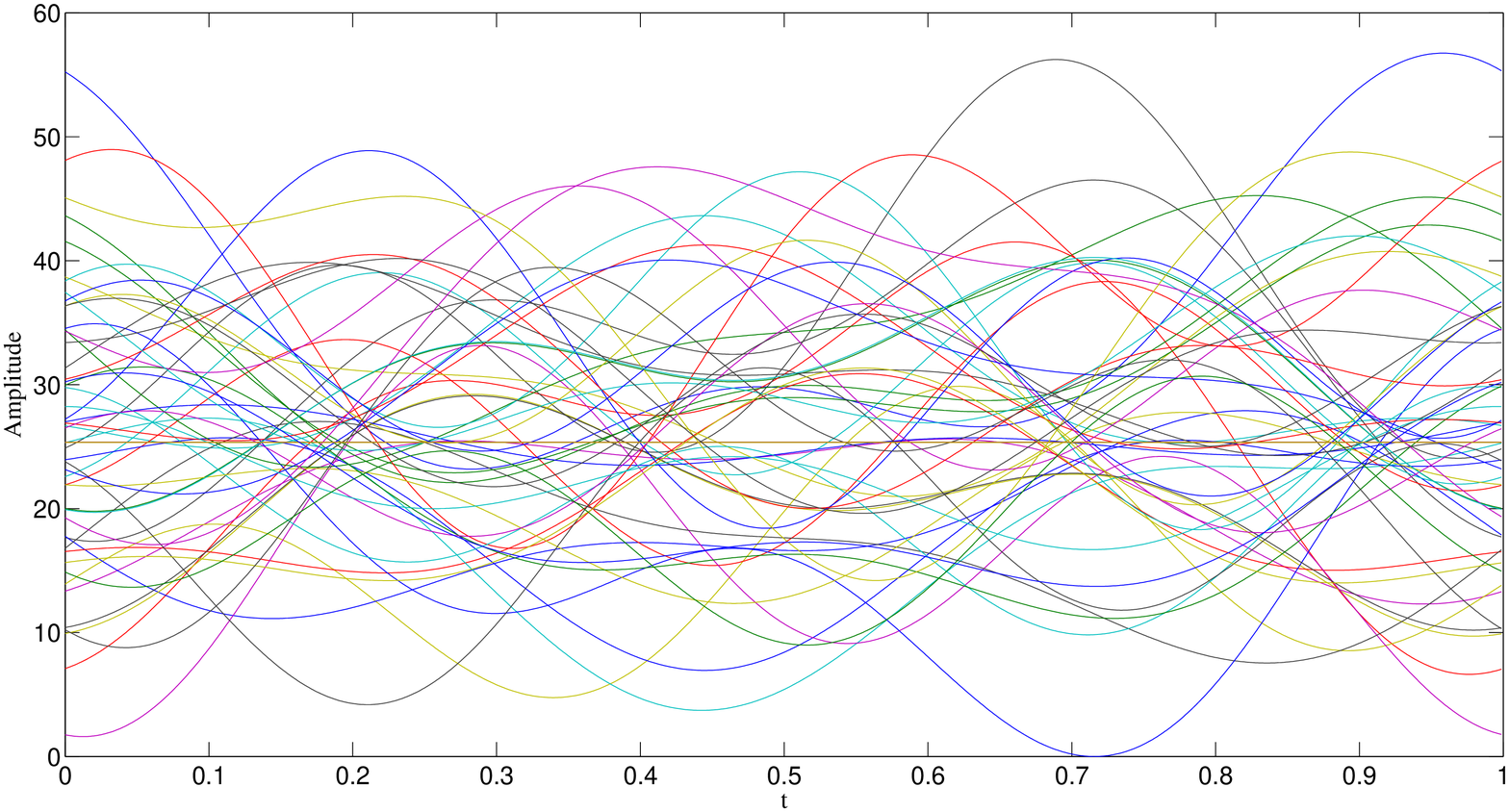}}
\vspace{-0.1in}
\caption{64 blue sub waveforms for DCI-JCFM with $K=2$, $\epsilon=0.1$.}
\end{figure}

\section{Conclusion}
We have propose a joint constellation design scheme termed DCI-JCFM taking advantage of the wavelength, frequency, and adaptive bias diversities at the same time for indoor visible light communication systems. By applying the DCI-JCFM scheme, waveform symbols with a much larger MED can be obtained than those from a decoupled scheme with or without channel cross-talks. Future works will include a comprehensive comparison among three systems: one applying dynamic range, one with long-term PAPR, and one with instantaneous PAPR constraint respectively; comparison of power efficiency of the DCI-JCFM and the popular DCO/ACO-OFDM schemes for multi-carrier multi-color VLC systems; and advanced precoder design to replace the SVD-based pre and post-equalizers utilized in this paper.

\end{document}